\documentclass[pra,twocolumn,superscriptaddress]{revtex4-1}

\usepackage{algorithm,algcompatible}
\newcommand{\INITIAL}{\item[\algorithmicinitial]}
\newcommand{\algorithmicinitial}{\textbf{Initial:}}
\newcommand{\RETURN}{\item[\textbf{Return:}]}

\usepackage{threeparttable}
\usepackage{diagbox}
\usepackage{tabularx}
\usepackage{amssymb}
\usepackage{graphicx}
\usepackage{dcolumn}
\usepackage{bm}
\usepackage{amsmath}
\usepackage{mathrsfs}
\usepackage{physics}
\usepackage{bbold}
\usepackage{comment}
\usepackage{color} 
\usepackage{textcomp}
\usepackage[unicode=true,bookmarks=true,bookmarksnumbered=false,bookmarksopen=false,breaklinks=false,pdfborder={0 0 1},backref=false,colorlinks=true]{hyperref}

\def\be{\begin{equation}}
\def\ee{\end{equation}}
\def\bea{\begin{eqnarray}}
\def\eea{\end{eqnarray}}

\begin{document}
\title{Quantum Adiabatic Algorithm Design using Reinforcement Learning}
\author{Jian Lin}
%\email{xiaopeng\_li@fudan.edu.cn}
\affiliation{State Key Laboratory of Surface Physics, and Department of Physics, Fudan University, Shanghai 200433, China}
\affiliation{Institute of Nanoelectronics and Quantum Computing,  Fudan University, Shanghai 200433, China}
%\affiliation{Collaborative Innovation Center of Advanced Microstructures, Nanjing 210093, China}
%\affiliation{These authors contributed equally to this work.}
%\affiliation{Collaborative Innovation Center of Advanced Microstructures, Nanjing 210093, China}
\author{Zhong Yuan Lai}
%\email{abrikosoff@gmail.com}
\affiliation{State Key Laboratory of Surface Physics, and Department of Physics, Fudan University, Shanghai 200433, China}
\affiliation{Institute of Nanoelectronics and Quantum Computing,  Fudan University, Shanghai 200433, China}
%\affiliation{Collaborative Innovation Center of Advanced Microstructures, Nanjing 210093, China}
%\affiliation{Collaborative Innovation Center of Advanced Microstructures, Nanjing 210093, China}
%\affiliation{These authors contributed equally to this work.}
\author{Xiaopeng Li}
\email{xiaopeng\_li@fudan.edu.cn}
%\affiliation{State Key Laboratory of Surface Physics, Institute of Nanoelectronics and Quantum Computing, and Department of Physics, Fudan University, Shanghai 200433, China}
\affiliation{State Key Laboratory of Surface Physics, and Department of Physics, Fudan University, Shanghai 200433, China}
\affiliation{Institute of Nanoelectronics and Quantum Computing,  Fudan University, Shanghai 200433, China}
%\affiliation{Department of Physics, Harvard University, Cambridge, Massachusetts 02138, USA} 
\affiliation{Collaborative Innovation Center of Advanced Microstructures, Nanjing 210093, China} 
\affiliation{Shanghai Research Center for Quantum Sciences, Shanghai 201315, China}

\begin{abstract}

{
Quantum algorithm design plays a crucial role in exploiting the computational advantage of quantum devices. 
 Here we develop a deep-reinforcement-learning based approach for quantum adiabatic algorithm design. Our approach is generically applicable to a class of problems with solution hard-to-find but easy-to-verify, e.g., searching and NP-complete problems. We benchmark this approach in Grover-search and 3-SAT problems, and find that the adiabatic-algorithm obtained by our RL approach leads to significant improvement in the resultant success probability.  In application to Grover search, our RL-design automatically produces an adiabatic quantum algorithm that has the quadratic speedup. 
 We find for all our studied cases that quantitatively the RL-designed algorithm has a better performance
 compared to the analytically constructed non-linear Hamiltonian path when the encoding Hamiltonian is solvable, and that this RL-design approach remains applicable even when the non-linear Hamiltonian path  is not analytically available. In 3-SAT, we find RL-design has fascinating transferability---the adiabatic algorithm obtained by training on a specific choice of clause number leads to better  performance consistently over the linear algorithm on different clause numbers. 
 These findings suggest the applicability of reinforcement learning for automated quantum adiabatic algorithm design. 
 Further considering the established complexity-equivalence of circuit and adiabatic quantum algorithms, we expect the RL-designed adiabatic algorithm to inspire novel circuit algorithms as well. Our approach is potentially applicable to different quantum hardwares from trapped-ions and optical-lattices to superconducting-qubit devices.
}

\end{abstract}

\date{\today}

%\pacs{67.85.-d, 03.75.Mn, 05.30.Jp, 05.30.Rt}

\maketitle

%\ncsection{Introduction}

\section{Introduction}
Quantum simulation and quantum computing have received enormous efforts in the last two decades owing to their  advantageous computational power over classical machines~\cite{preskill2012quantum,harrow2017quantum,lund2017quantum,bloch2018quantum,2019_Google_QuantumSupremacy}.  In the development of quantum computing, quantum algorithms with exponential speedups have long been providing driving forces for the field to advance, with the best known example from factorizing a large composite integer~\cite{shor1999polynomial}.  In applications of quantum advantage to generic computational problems, quantum algorithm design plays a central role.  In recent years, both threads of gate-based~\cite{nielsen2002quantum} and adiabatic annealing models~\cite{farhi_quantum_2000,2018_Lidar_RMP} of quantum computing have witnessed rapid progress in hardware developments such as superconducting~\cite{devoret2013superconducting,otterbach2017unsupervised,IBM,king2018observation,Google2018blueprint,2018_Pan_arXiv}, photonic~\cite{flamini2018photonic,brod2019photonic,wang2019boson}  
%~\cite{2017_Pan_NatPhoton,2017_White_PRL,2018_Jin_NSR,flamini2018photonic,brod2019photonic,wang2019boson} 
and atomic~\cite{2016_Monroe_NPJ,2017_Lukin_Nature,2018_Weiss_arXiv} quantum devices. Computational complexity equivalence between the two approaches have been established in theory~\cite{2001_Dam_arXiv,2008_Aharonov_Review,2018_Wu_CPL}.  

\begin{figure}[htp]
\includegraphics[width=.48\textwidth]{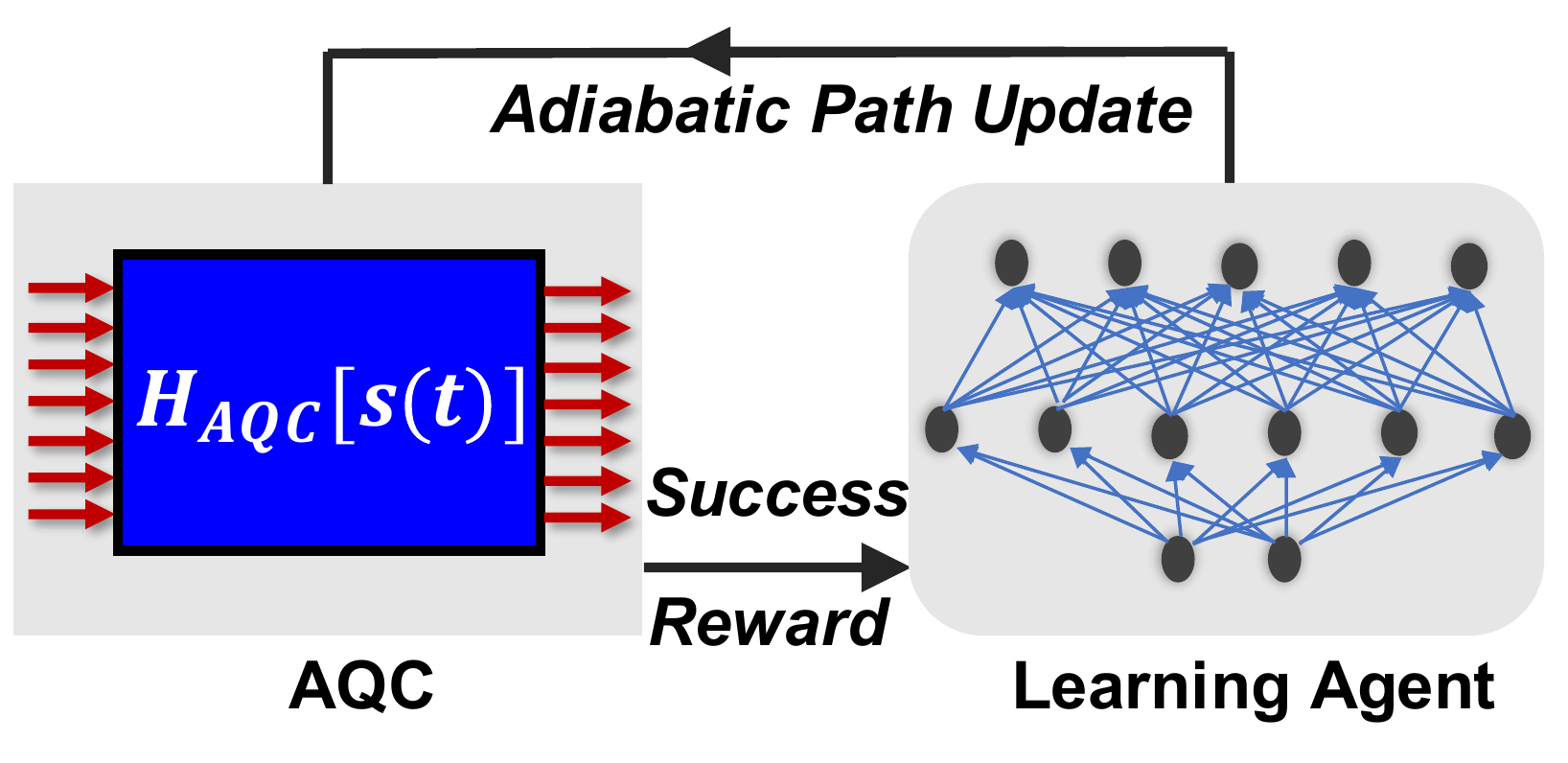}
\caption{Schematic illustration of the reinforcement learning (RL) approach for adiabatic quantum algorithm design. 
The RL agent collects a reward when the adiabatic quantum computer (AQC) finds the correct solution, whose efficiency relies on the solution being easy-to-verify. 
%This architecture is efficient for problems with solution hard-to-solve but easy-to-verify. 
%The RL agent takes the negative of the final quantum state energy of the adiabatic quantum computer (AQC) as a reward. 
The agent produces an action of adiabatic-path-update of $s(t)$ to optimize the reward based on its Q-table represented by a neural network 
{(see main text).} 
} 
\label{fig:schematic}
\end{figure}

In adiabatic quantum computing, the Hamiltonian can be written as a time-dependent combination of initial and final Hamiltonians, $H_B$ and $H_P$~\cite{farhi_quantum_2000,2018_Lidar_RMP}, as 
\be
\textstyle H = (1-s(t/T))H_B + s(t/T)H_P , 
\label{eq:Ham} 
\ee 
with the computational problem encoded in the ground state of $H_P$. In this framework, the quantum algorithm design corresponds to the optimization of the Hamiltonian path or more explicitly the time sequence of $s(t)$. Different choices for the path could lead to algorithms having dramatically different performance and even  in the complexity scaling. For example in Grover search, a linear function of $s(t/T)$ leads to an algorithm with a linear complexity scaling to the search space dimension ($N$), whereas a nonlinear choice could reduce the complexity to $\sqrt{N}$~\cite{roland_quantum_2002}. This   implies an approach of automated quantum adiabatic algorithm design through searching for an optimal Hamiltonian path, which may lead to a generic approach of automated algorithm design given the established complexity equivalence between gate-based and adiabatic models~\cite{2001_Dam_arXiv,2008_Aharonov_Review,2018_Wu_CPL}. The automated quantum algorithm design that is adaptable to moderate-qubit-numbers is particularly in current-demand considering near term applications of  noisy intermediate size quantum devices~\cite{2018_Preskill_NISQ}.

Here, we propose a deep reinforcement learning (RL) architecture for automated design of quantum adiabatic algorithm. By encoding the computation problem in a Hamiltonian ground state problem, we find that the automated design of quantum algorithm can be reached by RL of the optimal Hamiltonian path. 
Our RL architecture is most efficient to a class of problems with solutions easy-to-verify, e.g., searching, factorization, and NP-complete problems. 
In application to the Grover search and 3-SAT problems, we find that the adiabatic algorithm designed by the machine has a better performance than linear algorithms in terms of computing efficiency or the success probability.

For the Grover search, the RL-design automatically produces an adiabatic algorithm that takes as much time as the nonlinear path~\cite{roland_quantum_2002} when the Grover search Hamiltonian is solvable and the solution of nonlinear path is analytically available. And quantitatively, with the same amount of time the RL-designed algorithm reaches a higher success probability than the analytically constructed non-linear path. 
When the analytical nonlinear path is unavailable in using a non-solvable Grover search Hamiltonian encoding, we still find  that RL-design 
 produces an adiabatic algorithm whose time resources scale as $\sqrt{N}$.
 
 For the 3-SAT problem,   the RL-designed quantum algorithm is found to have  emergent transferability---the algorithm obtained by  training on a subset of problem instances is applicable to other very different ones while maintaining the  high computational performance. 
This transferability is not only conceptually novel but also practically crucial in saving computation resources for training. 

%\medskip 
%\ncsection{Results}      
\section{Reinforcement learning architecture for quantum adiabatic algorithm design}
\subsection{Adiabatic algorithm design as an optimization problem}
\label{sec:IIA} 
%{\it Reinforcement learning architecture for quantum adiabatic algorithm design.---} 
Given a computational problem, e.g., Grover search or 3-SAT, the form of the  Hamiltonian $H_P$ encoding the problem is fixed.  For different problem instances, for example in targeting different states in Grover search or finding solutions for different choices of clauses in 3-SAT, the encoding Hamiltonian is different. We label different  problem instances by $PI$, and the encoding Hamiltonian is correspondingly labeled as $H_{PI}$. The designed Hamiltonian path in general would depend on the computational  problem, for example  whether it is Grover search or 3-SAT, but it should be required that the Hamiltonian path should be independent of the problem instance $PI$, in order for this Hamiltonian path design to make a   quantum adiabatic algorithm generically applicable. This makes it distinct from path optimization aiming for preparation of specific quantum states~\cite{2018_Bukov_PRX} or for  achieving robust or fast gate operations~\cite{berry2009transitionless,chen2010shortcut,2018_Wang_PRA,niu2018universal}.   

We propose an approach for automated algorithm design based on reinforcement learning (see Fig.~\ref{fig:schematic} for an illustration). In the framework of quantum  adiabatic algorithm, the task of algorithm design reduces to the  exploration of the optimal path $s(t/T)$, which we parameterize  as 
\be 
\textstyle s\left(\frac{t}{T}\right) = \frac{t}{T} + \sum_ {m=1} ^{C} b _m  \sin(\frac{m\pi t}{T}). 
\ee   
Here $C$ is a cutoff for high frequency components, and the parameters $b_m $ form a vector ${\bf b}$. This parametrization is asymptotically complete as the cutoff $C$ approaches infinity. 

To build an artificial intelligent agent that explores the path-space  of ${\bf b}$, we introduce a set of action, {\bf a}, which are defined to update ${\bf b}$ as 
$a^{(0)} ({\bf b}) = {\bf b}$ and  $[a^ {(2m-1)} ({\bf b})] _{n}  = b_{n}  - \Delta_0 \delta_{mn}$, $[a ^{(2m)} ({\bf b})] _{n} = b_{n} + \Delta_0 \delta_{mn}$ for $m\ge 1$, with $\Delta_0$ to be referred to as maximal update per step and the Kronecker delta $\delta_{mn}$.
%\bea 
%&& a_0 ({\bf b}) = {\bf b}   \\
%&& [a_m ^\pm ({\bf b}) ]_n = b_n \pm \Delta \delta_{mn} . \nn 
%\eea 

A unit reward, $r$, is collected by the agent if the solution out of  the adiabatic quantum computer is correct. 
This approach then directly applies to adiabatic quantum hardwares such as D-wave machines~\cite{Dwave} for the class of problems with solutions hard-to-solve but easy-to-verify. 
% is assigned to be the opposite of the final quantum state energy following the Hamiltonian (Eq.~\eqref{eq:Ham}) evolution given by ${\bf b}$.  
To target an optimal adiabatic algorithm with robust performance to all problem instances, we sample $PI$ and average over a certain number  of instances ($MI$)  in calculating the reward for an action ${a}$ on ${\bf b}$. 
In the reinforcement learning approach, during an intermediate $j$-th step, the agent evaluates the action ${a}$ on ${\bf b}$  according to a Q-table 
 $ 
 Q^\star ({\bf b}, {a})=\max_{\cal P} \mathbb{E}\left[ \sum_ {i=0} ^{\infty} \gamma^i r \left( {\bf b} ({j+i}) \right) |{\cal P} \right], 
$ 
where $\gamma \in (0,1)$ is a discount factor that allows to account for future rewards of ${\bf b} (j+i)$, and ${\cal P}$ represents an action-selecting policy describing the probability of performing the action ${a}$ on a path-state ${\bf b}$. 
%This  maximizes the expected cumulative future reward  by choosing an action-selecting policy ${\cal P}=P(a |{\bf b})$ that describes the probability of performing the action ${a}$ on a path-state ${\bf b}$. 

Following the action selection,  we use a protocol analogous to simulated annealing and update the path state stochastically according to 
a probability determined by the the corresponding action Q-value. The path state is updated in this way because our Hamiltonian path space is continuous, distinct from the Go-game where the reinforcement learning has already found a great success~\cite{2017_AlphaZero}. 
Details of the state-update policy are described in Sec.~\ref{sec:policy}. 

%In our method, although the selection of the action is probabilistic,  the next step path-state ${\bf b}_{j+1}$  is deterministically set by the $j$-th step action and path-state, ${a}_j$ and ${\bf b}_j$. % as ${\bf b}_{j+1} = {a}_j ({\bf b}_j)$. 
%In accounting for the future reward, the $(j+i)$-th step reward is discounted by a factor of $\gamma^i$. 
The Q-table  can be solved by iteration according to the Bellman equation~\cite{bellman1952theory}. 
%{
%$
% Q^*({\bf b}, a)= \left[ r ( a( {\bf b} )) +  \gamma\, {\rm max}_{a'} Q^*( a( {\bf b} ), { a}') \right]. 
%$ 
%}

\begin{figure*}[htp]
\includegraphics[width=.9\textwidth]{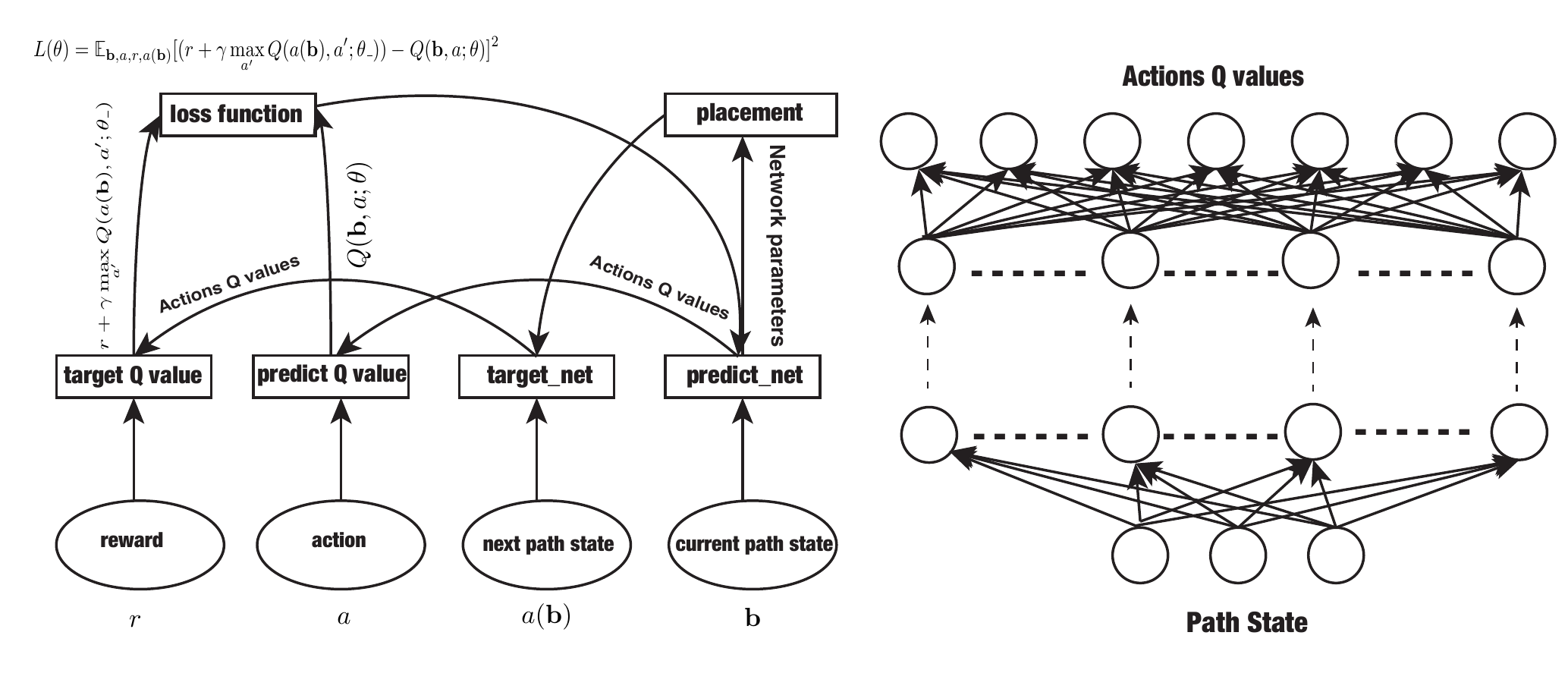}
\caption{Sketch of the RL architecture used in this work. 
Left panel shows the relationship between various quantities and functions defining the RL framework. The `predict\_net' and `target\_net' are two networks having network-parameters $\theta$ and $\theta_-$ (see Section~\ref{sec:IIA}  and algorithm~\ref{alg:architecture}), which represent the Q-table in the reinforcement learning. In the training process,  the network `predict\_net' is trained on the fly---it is updated in every training step. 
The network parameters in `target\_net' are updated to that in `predict\_net' every $W$ steps by `placement' as explained in the main text. 
We train the network `predict\_net' such that its produced Q-value through the function `predict Q value' matches the reward of the current path state plus the discounted maximal Q-value of actions performing on the next path state, produced by the delayed network `target\_net' and the function `target Q value'.    
The `loss function' is defined accordingly.    
%represent the Q-table of different actions according to the current path state ${\bf b}$ and the next path state ${a(\bf b)}$, respectively. 
%The action $a$ is selected according to the state-update policy (see Section~\ref{sec:policy}) and the reward $r$ is obtained by averaging the performance of $MI$ instances. We perform a stochastic gradient descent on the the loss function which depends on the target Q value and predict Q value with respect to the `predict\_net' parameters. And the `target\_net' network parameters update to `predict\_net' network parameters every $W$ steps 
%(see main text and algorithm~\ref{alg:architecture}). 
Right panel illustrates the structure of the fully-connected multi-layer neural networks representing the Q-table whose input is the path state and the output is the Q-values corresponding to the different actions.}
\label{fig:RL_NN}
\end{figure*}

Our method uses a deep neural network to approximate the Q-table, as $ Q^\star ({\bf b}, { a}) \approx Q({\bf b}, { a}; \theta)$, with $\theta$  the network parameters determined in an iterative learning process.  To stabilize the  nonlinear iteration in learning, 
 we adopt an experience-replay approach~\cite{mnih2015humanlevel} where the agent's experiences $({\bf b}, a,  r)$ are stored in a memory ${\cal M}$ with a capacity $CAP$.  
We have a network $Q({\bf b}, a; \theta)$ trained {\it on-the-fly} during  the agent exploring the path-state space. As for the training, the inputs and outputs are  
${\bf b}$ and  
%$r + \gamma\, {\rm max}_{a'} Q({\bf b}', a'; \theta_-) $, 
$r + \gamma\, {\rm max}_{a'} Q(a( {\bf b}), a'; \theta_-) $
respectively, 
with ${\bf b}$, $r$, and $a$ drawn randomly from the memory ${\cal M}$, 
%and $Q({\bf b}', a'; \theta_-)$ a separate network whose 
and the network parameter $\theta_-$  updated to $\theta$  every $W$ steps.

Given the limitations of quantum computing hardwares presently accessible, we simulate quantum computing on a classical computer and generate reward to train the RL network. In applications to a quantum computer,  our RL architecture for automated algorithm design is directly adaptable by collecting reward generated by a quantum adiabatic computer. 

\subsection{ Neural network architecture }
In Fig.~\ref{fig:RL_NN} we show the explicit learning protocol. The left-hand panel of Fig.~\ref{fig:RL_NN} shows a schematic of our RL framework. We use a two-network setup which is similar to that used in Ref.~\onlinecite{mnih2015humanlevel}. These two networks are labeled `target\_net' and `predict\_net' in the schematic. 
%They have exactly the same structure. 
The `target\_net' uses a previously-learned set of parameters and updates the parameters every $W$ steps; the values of these parameters are directly copied from `predict\_net'. This will reduce problems such as divergences and oscillations thus making the learning process more stable.
The loss function calculates the expectation value of the difference between `target Q value' and `predict Q value' over different batches. 
The right-hand panel shows the multi-layer fully-connected neural network used to represent the Q-table. The input to this network is the path state obtained at some learning step, while the output are the Q-values corresponding to the actions.

\subsection{ Details of the state-update policy in the reinforcement learning }
\label{sec:policy} 

For the state-update policy, we used the $\epsilon$ greedy strategy, where the agent selects a random action with total probability $1-\epsilon$, and the action having the maximal Q-value with an additional probability $\epsilon$.   We set the $\epsilon=0$ initially and let  the agent explore the state space with no preference. The value of $\epsilon$ is increased gradually until a maximum value $90\%$  in the learning process. In this way, we try to maintain a balance between the agent's current knowledge to maximize reward and possibilities in exploring among other options, which in turn leads to a learning process with better performance.

%Depending on the value of $\epsilon$ this strategy might take a long time to converge to the global minimum which signifies the solution to the problem. 
In order to deal with the continuous state-space, we develop a continuous state-update protocol and combine with the $\epsilon$-greedy method in the following way. 
%In order to improve convergence of the continues moving RL algorithm we introduce a modified~\emph{simulated annealing} procedure in addition to the greedy strategy. 
In our procedure, we use the $\epsilon$-greedy strategy %only~\emph{after} a certain threshold for the reward has been reached; before reaching this threshold,
to choose an action. After choosing the action, the agent sets a probability to accept the action which 
 we calculate from an~\emph{acceptance probability function} $P(e, Tem) = \exp\left(\frac{e}{Tem}\right)$, where
\be\label{eq:energy_diff}
e = \frac{Q(a(\mathbf{b}),a;\theta) - Q(\mathbf{b},a;\theta)}{\Delta_0} * \Delta,\,\,\,\,\,\,\Delta\in [0, \Delta_0],
\ee
$Tem$ is a modifiable parameter commonly identified as the ``temperature'' of the system in the context of simulated annealing methods, 
%and $Q$ is the output of the deep neural network defined in the Methods section. 
The ``temperature" is annealed down every agent-exploration step (labeled by $j$), following $Tem_j=Tem_{j-1}\times10^{-C_R}$, withe $C_R$ the cooling rate. 
%The effect of $P(e,T)$ can be better understood in context of the annealing procedure, which we outline below.
  We run cycles of this $\epsilon$-greedy learning process to ensure convergence of the Q-table. 
%which is analogous to the reannealing approach in simulated annealing method. 
At each step of RL exploration, the neural network is trained by varying the  $\theta$ parameters to solve an iteration problem, 
\be 
Q({\bf b}, a; \theta) = r(a({\bf b})) + \gamma  \,{\rm max}_{a'} \left[ Q(a({\bf b}), a'; \theta_-) \right]. 
\ee 
The training data is generated from the memory ${\cal M}$ that stores the path-states ${\bf b}$, actions $a$, and the corresponding rewards $r(a ({\bf b}))$ that the RL agent has explored. 
The parameters $\theta_-$ are only updated to $\theta$ every $W$ steps ($W$ is set to be $50$ here), to  deliberately slow down the iteration process for stabilization purpose. 
This approach has been used in Ref.~\onlinecite{mnih2015humanlevel}, and follows  a standard approach to stabilize nonlinear iteration problems.   When the iteration converges, $Q({\bf b}, a; \theta)$ satisfies the Bellman equation~\cite{bellman1952theory}.

The effect of $P(e,Tem)$ can be better understood in context of the annealing procedure, which we outline below.
At intermediate step $j$ the agent evaluates the action $a_j$ on the path state $\mathbf{b}_j$, which then defines the path state $\mathbf{b}_{j+1}$ and reward at step $j+1$. At this point we perform the following steps, 
\begin{itemize}
    \item Fix $\Delta_0$ to some constant; 
    \item Calculate~\eqref{eq:energy_diff} with values of the parameters as obtained at step $j$;
    \item Generate a random number $\mu\in[0.0, 1.0]$
    %\begin{itemize}
    \item If $\mu\leq\exp\left(\frac{e}{Tem}\right)$
         \begin{itemize}
             \item Accept corresponding action and accept the current value of $\Delta$.
         \end{itemize}
         \item else choose action $= 0$ (corresponding to taking no action).
    %\end{itemize}
\end{itemize}
%At this point we store the current path state, action taken, reward resulting from action on path state and the corresponding next path state. We measure the value of the fidelity with respect to the new path state. If the value of the fidelity is larger than $0.999$ the algorithm then tries to accelerate convergence even further by reverting to the $\varepsilon$-greedy strategy and setting $\varepsilon_{max}$ to $1$. We then iterate the above steps for the new path state and parameter values.
In the learning process, we store the current path, action taken, reward resulting from action on path state and the corresponding next path state. 
%\textbf{Here we set a threshold of fidelity $99.9\%$ and when the average reward reaches the threshold} 

When the reward reaches a threshold of $99.9\%$---a maximal reward is $1$ in our notation, as it resembles the success probability of the RL-designed adiabatic algorithm, 
the value of $\varepsilon_{max}$  is set to $1$, and the agent then uses the network to choose action.  
In exploring the Hamiltonian path-state space, as the iteration step increases, it gets more frequent for the agent to 
find a path that gives a higher success probability.
%The convergence of the training process is confirmed in our calculation. 

After the Q-table converges, we let the agent update the path-state ${\bf b}$ until it stabilizes, according to the $\epsilon$-greedy policy with $\epsilon$ increased from $90\%$ to $100 \%$ slowly with fixed annealing temperature and neural network parameters

\begin{figure}[htp]
\includegraphics[width=.45\textwidth]{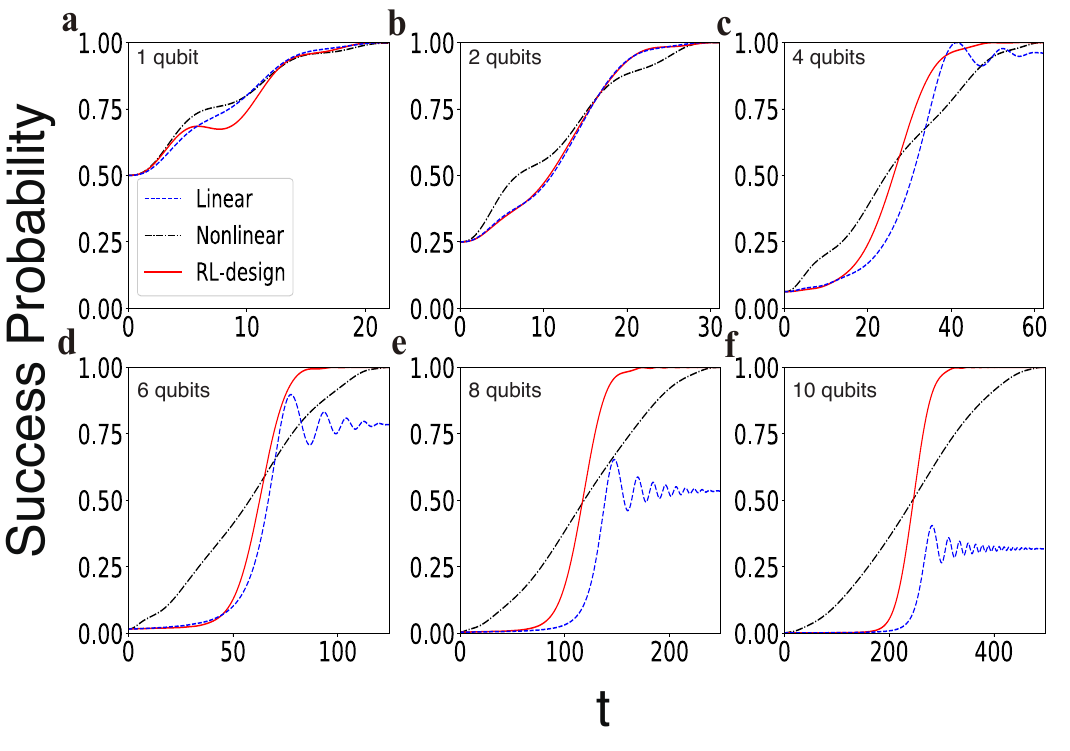}
\caption{Performance of RL-designed quantum adiabatic algorithm in success probability for Grover search.   The success probability is obtained by taking the square of wave-function overlap of the dynamical quantum state with search-target state. Results from adiabatic algorithms using  a linear and a tailored nonlinear path~\cite{roland_quantum_2002} are shown for comparison. The total adiabatic time are chosen to be {$T = 22.0, 31.1, 62.2, 124.5, 248.9, 497.8$} 
for qubit number $n = 1, 2, 4, 6, 8, 10$,  respectively, following the $\sqrt{N} = \sqrt{2^n}$ scaling.
 Given this choice of scaling, an eventual success probability close to $1$ by both of the non-linear  and the RL-designed algorithms implies that they both exhibit quadratic quantum speedup because otherwise the success probability would dropdown with increasing qubit number. 
Comparing the RL-designed and non-linear algorithms quantitatively, it takes less amount of time for RL-designed algorithm to converge to the searching target than the non-linear algorithm.}

%The machinery  adiabatic algorithm designed by RL shows significant improvement over the linear algorithm, and reveals the same computation-complexity scaling as the nonlinear algorithm. 

\label{fig:fidelity}
\end{figure}

%\medskip 
\section{Performance on Grover search}
%{\it Performance in Grover search.---} 
\subsection{ Learning of easy Grover search} 

\begin{figure*}[htp]
        \includegraphics[width=.8\textwidth]{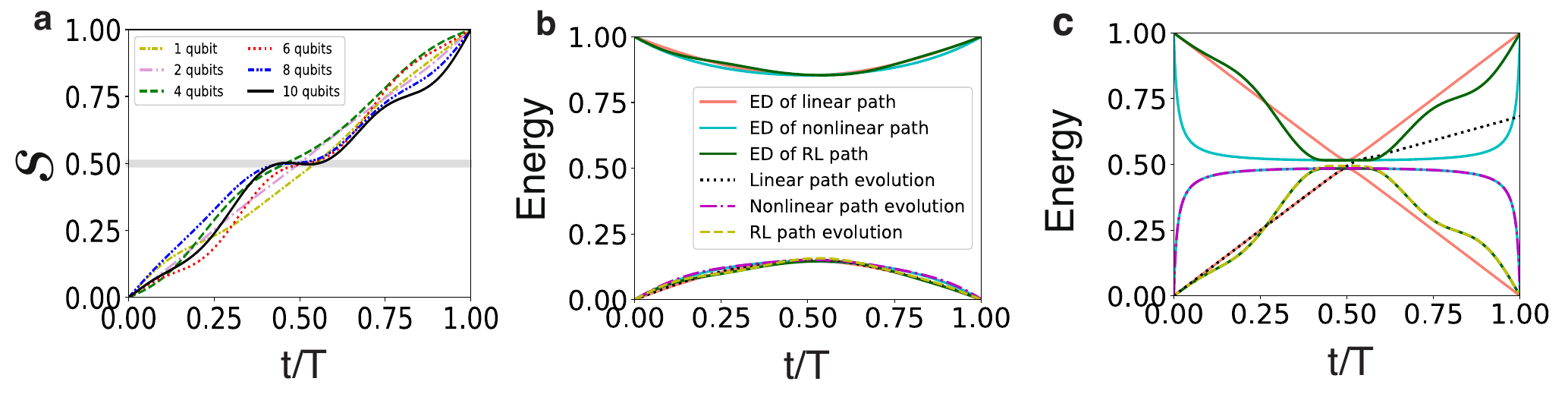}
  \caption{Energy evolution from the RL-designed adiabatic path for Grover search. ({\bf a}) shows the RL-designed path. The adiabatic time is chosen the same way as in Fig.~2 ({\it see the main text}). ({\bf b}) and  ({\bf c}) show the energy spectrum for the ground and first excited states with $1$ and $10$ qubits, respectively. The energy spectra of the instantaneous Hamiltonian are obtained by exact diagonalization (ED), shown by solid lines in ({\bf b}, {\bf c}). The plot in ({\bf c}) shares the same legend as in ({\bf b}).  
  The energy expectation values of the  dynamical state following different Hamiltonian paths are shown by dashed lines.  It is evident from (${\bf c}$) that the RL-designed path is distinct from both of the linear and the nonlinear paths. 
} 
%\endminipage
\label{fig:combined_fig}
\end{figure*}

In application of our RL approach to automated adiabatic algorithm design, we first show its performance on Grover search compared to known quantum algorithms. This search problem is to find an element in an array of length $N$ as an input to a black-box function that produces a particular output value. This classical problem can be encoded as searching in the Hilbert space of $n=\log_2 N$ qubits for a target quantum state. These qubits are labeled by $q$ in the following. A circuit-based quantum algorithm was firstly designed by Grover, which shows a quadratic quantum speedup over classical computing~\cite{grover1996fast}. In adiabatic quantum  computing, the Hamiltonians in Eq.~\eqref{eq:Ham} for Grover search are 
$H_B =\mathbb{1}-|\psi_0\rangle \langle \psi_0|$, and $H_P = \mathbb{1}-|m\rangle\langle m| $, where  $|m\rangle$ is a product state in Pauli-$Z$ basis that encodes the search target, and $|\psi_0\rangle$ is a product state in the Pauli-$X$ basis with  all $n$ eigenvalues equal to $1$. The symbols $X$, $Y$, and $Z$ refer to Pauli matrices in this work. A linear choice of $s(t/T)$ (${\bf b} = 0$ in our notation), does not exhibit the quadratic  speedup. It was later pointed out in Ref.~\onlinecite{roland_quantum_2002}  that the quantum speedup  is restored with a tailored nonlinear path choice of $s(t/T)$. 

%It is worth remarking here that despite the knonwn complexity scaling comparison, it is unclear what path choice is optimal for moderate Grove search problems. 

%Hamilonian. 
%Emphasize more about  the algorithm. 
%This problem actually does not require sampling different problem instances. 

{
In the Grover search problem, different problem instances correspond to different choices for the $|m\rangle$ states, which are all connected to each other by a unitary transformation 
%$\otimes_{\{q\}} X_{\{q\}}$ for a certain subset of qubits $\{q\}$, 
which keeps $H_B$ invariant. The reward RL-agent collects in the training process is thus exactly equivalent for different problem instances, which means averaging over $PI$ is unnecessary for the Grover search. 
Fig.~\ref{fig:fidelity}  shows results of the RL-designed adiabatic Grover search algorithm. 
In our RL design for adiabatic quantum algorithm, we scale up the adiabatic time $T$ as $T\propto \sqrt{N}$ to benchmark against the best-known Grover search algorithm. 
Then as expected, the linear adiabatic algorithm leads to a success probability completely unsatisfactory at large $N$. 
We find that both the nonlinear~\cite{roland_quantum_2002} and the RL-designed adiabatic algorithms produce success probabilities very close to $1$ (larger than {$99.9\%$)}.  At large $N\ge 2^4$, the RL-designed algorithm outperforms the nonlinear one.

In Fig.~\ref{fig:fidelity}, a quantitative comparison shows that the required adiabatic time $T$ to  reach a  fixed  success probability is shorter from the RL-designed algorithm than the analytically constructed non-linear algorithm. 
We want to emphasize here that in Fig.~\ref{fig:fidelity} we scale up the total adiabatic time according to the $\sqrt{N} = \sqrt{2^n}$ scaling. Having an eventual success probability close to $1$ implies the the computational complexity of the RL-designed adiabatic algorithm follows the $\sqrt{N}$ scaling, because otherwise the success probability would significantly decrease as we increase the qubit number from $1$ to $10$. 
%The computational complexity of the RL-designed adiabatic algorithm is thus expected to follow  the $\sqrt{N}$ scaling, 
The $\sqrt{N}$ scaling is already known to be optimal for Grover search~\cite{1998_Farhi_PRA}.

{
It is worth remarking that  the choice of $H_B$ is made here for comparison purposes, as the nonlinear path to achieve the quadratic speedup is only analytically available with that specific Hamiltonian choice~\cite{roland_quantum_2002}. For physical realization of $H_B$, which can be rewritten as  $ H_B = \mathbb{1}-\otimes_q [ 1+X_q]/2$, it is experimentally challenging to construct this Hamiltonian with quantum annealing devices. A more suitable choice for $H_B$ in that regard  is $\sum_q [\mathbb{1}-X_q]/2$, for which  the analytically obtained nonlinear path~\cite{roland_quantum_2002} is then no longer applicable, but our RL design still produces high-performance adiabatic algorithms. 
A common feature of the RL-learned path for $s(t/T)$ is that there is a relatively flat region around $s = 0.5$ where the energy gap is minimal. This flat region has a tendency to grow as  we increase $N$ (Fig.~\ref{fig:combined_fig}(a)). 
}

%(See Supplement).    

%We also emphasize here that our RL-design does not lead to shortcut-to-adiabatic passage for two reasons---(i) the path is 
%However the case becomes different as we increase the qubit number. Fig.~\ref{fig:combined_fig} (c) shows the energy evolution for ten qubits. We find that the instaneous dynamical quantum  states for the  RL-designed path follow the instaneous ground state of the Hamiltonian, and that  with the adiabatic time $T = 2500$, the difference between the energy of the instaneous dynamical quantum states  and the ground state energy is barefly noticable for the RL-designed path, much smaller than the linear path. 

%We expect our results to inspire circuit quantum algorithm design with better practical performance on NISQ as well. 
\subsection{Instantaneous energy spectrum for the easy Grover search} 

In this section, we show the resultant instantaneous energy spectrum corresponding to the RL-designed algorithm for the easy-way Grover search.
% A common feature of the RL-learned path for $s(t/T)$ is that there is a relatively flat region around $s = 0.5$ where the energy gap is minimal. This flat region has a tendency to grow as  we increase $N$ . 

%When the qubit number is small (Fig.~\ref{fig:combined_fig}(b)), the instaneous energy of the dynamical states for both the linear and RL-designed path deviates from the actual instaneous ground state. The difference is that the energy of  dynamical state from the RL-designed path bends more towards the targeted ground state when $t/T$ approaches $1$. This implies that the RL-designed path for small number of qubits is more like shortcut passage to adiabaticity than could be taken as an adiabatic algorithm. 
The instantaneous energy following the RL-designed algorithm lies on   the time-dependent ground state for both small and large number of qubits (Fig.~\ref{fig:combined_fig}(b,c)). The energy deviation is much smaller than the linear algorithm, and is very close to the tailored nonlinear algorithm. Therefore the RL-design approach  indeed automatically reveals a  quantum adiabatic algorithm as efficient as the improved nonlinear algorithm for Grover search~\cite{roland_quantum_2002}.

\begin{figure}[htp]

\includegraphics[width=.5\textwidth]{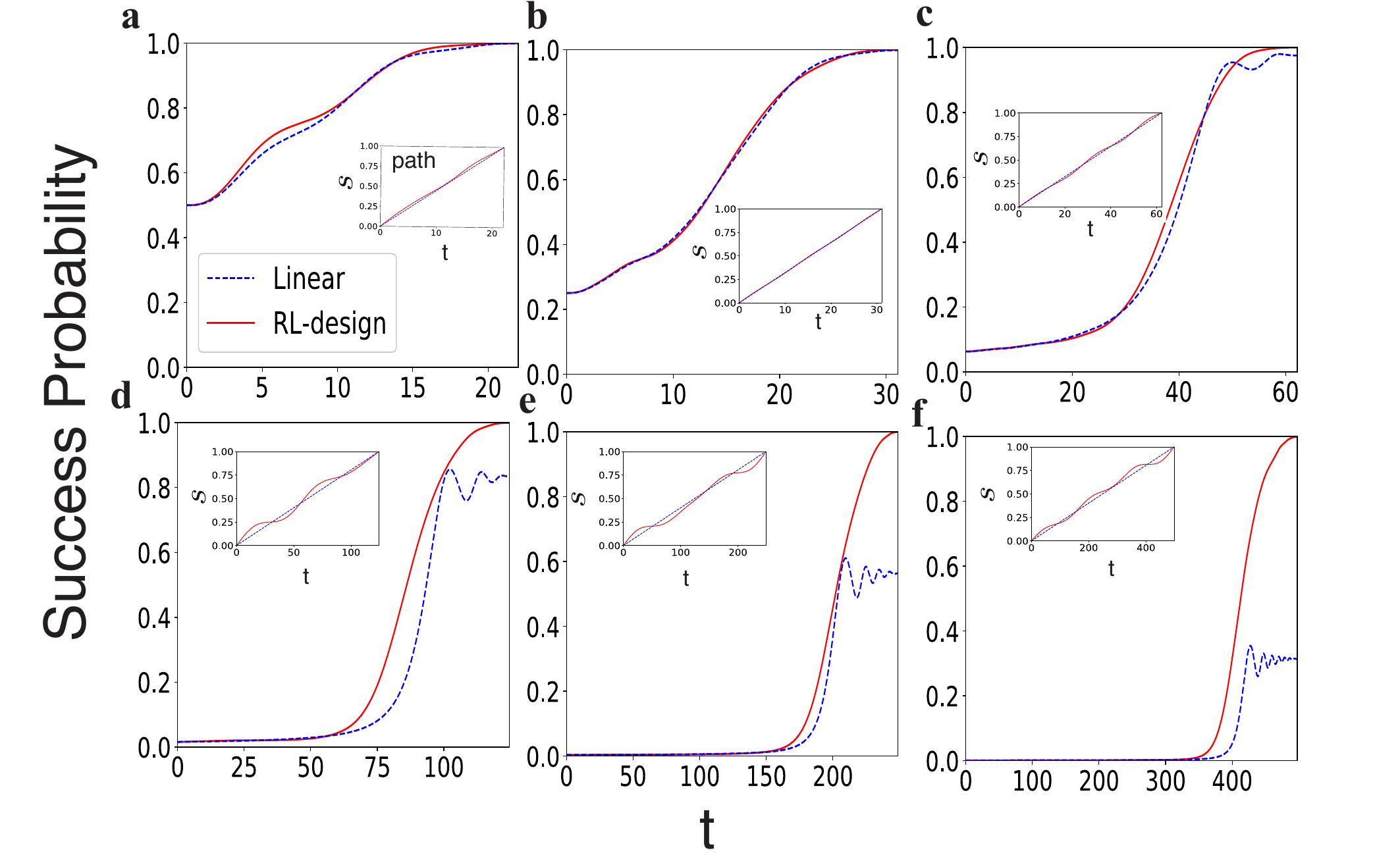}
\caption{Success probabilities for the hard Grover problem for different number of qubits. Comparison is between a linear protocol and our RL-designed protocol. The nonlinear protocol is not applicable here. The qubit numbers are, from top to bottom, left to right, $n = 1, 2, 4, 6, 8, 10$ for the the adiabatic evolution times $T = 22.0, 31.1, 62.2, 124.5, 248.9, 497.8$.}
\label{fig:paths_HG}
\end{figure}

\begin{figure}[htp]
\includegraphics[width=.48\textwidth]{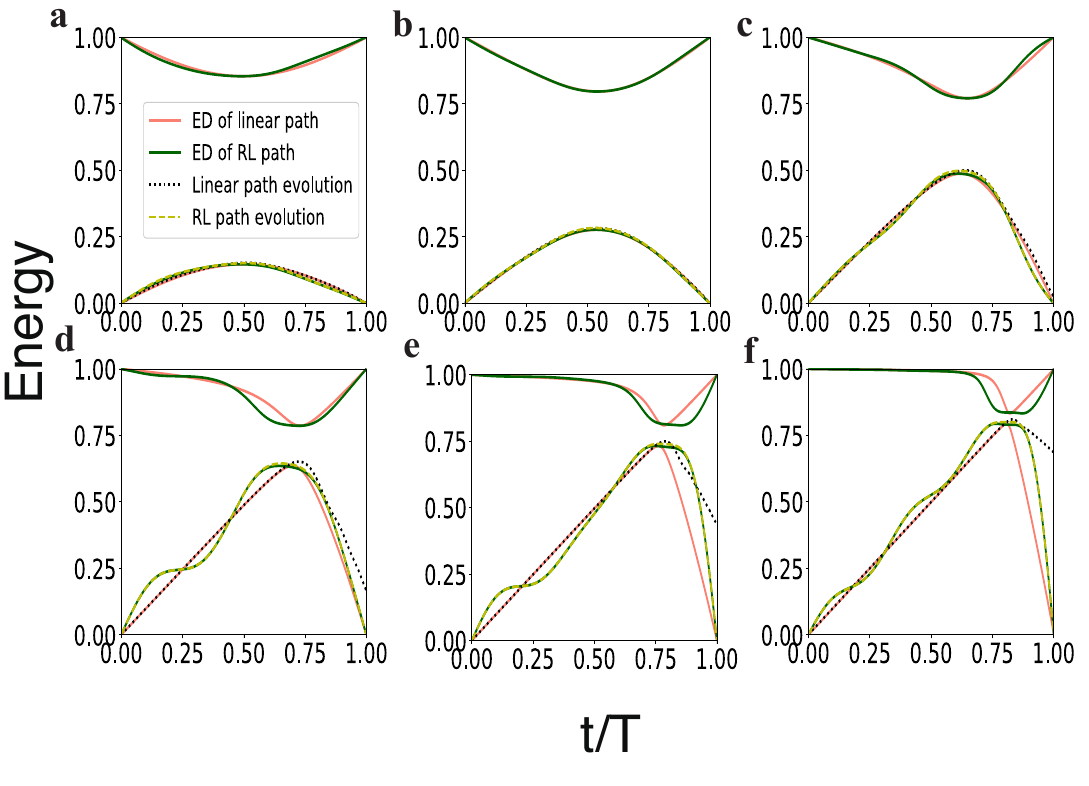}
\caption{Energy spectrum for the hard Grover problem for different number of qubits. The qubit numbers $n$ and adiabatic evolution times $T$ are the same as those given in the caption of Fig.~\ref{fig:paths_HG}. The expectation value of the instantaneous Hamiltonian by ED and the dynamical state following linear and RL-designed Hamiltonian path are shown by `solid' and `dashed' line. We do not show the nonlinear results as they are not applicable here.}
\label{fig:spectrum_HG}
\end{figure}

\subsection{Hard Grover search}

In learning the adiabatic algorithm of the Grover search, we choose the analytical-solvable Hamiltonian as in Ref.~\onlinecite{roland_quantum_2002} for comparison purpose, where the quantum dynamics during the adiabatic procedure corresponds to an effective two-level system. 
%the results concerning the Grover search problem using a simplified version with the special initial Hamiltonian $H_B$,
%\be
%H_B = 1 - |\psi_0\rangle\langle \psi_0|
%\ee
%where $|\psi_0\rangle = \frac{1}{\sqrt N}\sum_{i=0}^{N-1}|i\rangle$, and $N$ is the Hilbert space dimension. Because the special version can be considered as a physical two-level system, 

We  thus denote this as the ``easy'' Grover problem. Considering  physical realization,  a suitable choice for the encoding Hamiltonian $H_B$ is 
\be
H_B = \sum_q[\mathbb{1}-X_q]/2
\ee
We denote this the ``hard'' Grover problem, for which the analytically obtained nonlinear path~\cite{roland_quantum_2002} does not carry over~\cite{2018_Jarret_Wan_arXiv,2019_Slutskii_arXiv}. 
We stress here that  our RL approach still produces an adiabatic algorithm with high success probability. 
In this regard, our RL approach is more generic, and is particularly useful considering the present limitations of quantum hardwares. 
%The Hilbert space dimension will increase as the qubit number increases and the complexity of the hard problem is therefore much higher than that of the easy problem.

\begin{figure*}[htp]
\includegraphics[width=.8\textwidth]{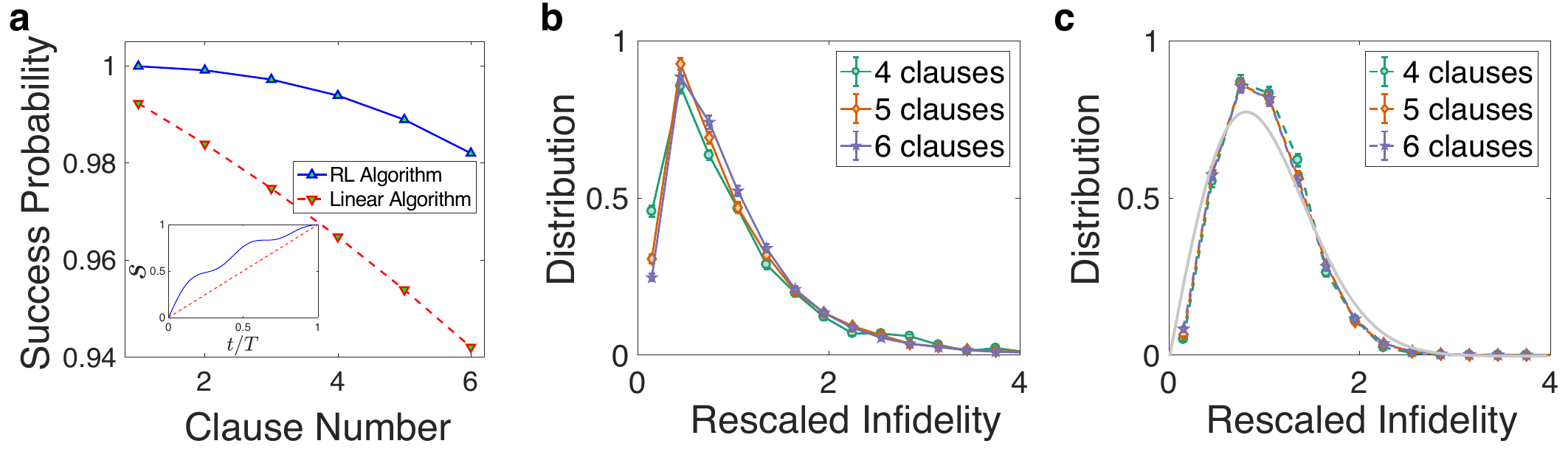}
\caption{Performance of reinforcement learning (RL) designed algorithm on 3-SAT problem. 
In ({\bf a}), we show the comparison of  the RL-designed and linear algorithms in the averaged success probability of solving random 3-SAT problems.  
The success probability is defined by projecting the final state of the adiabatic quantum evolution onto the correct solutions. ({\bf b}) and ({\bf c}) show the distribution of the infidelity for the RL-designed and linear algorithms, respectively. The statistics is collected from solving $10^5$ random 3-SAT problem instances using the corresponding quantum algorithms. The error bar represents the statistical error according to the bootstrap method. The infidelity is rescaled by taking its average as a unit. The infidelity distribution from the linear algorithm shown in (${\bf c}$) empirically resembles a Wigner-Dyson type 
(illustrated by the `grey' solid line), whereas the distribution from the RL algorithm in (${\bf b}$) deviates from that.  In this plot we choose the adiabatic time $T=6$, and the total bit number $N_b = 10$.
}
\label{fig:3fig_combine}
\end{figure*}

In Fig.~\ref{fig:paths_HG}, we compare results obtained from the linear protocol with those obtained from the RL-learned protocol for the hard Grover problem. We show the success probabilities as obtained from the linear and RL designed path for different numbers of qubits. Similarly to our results in Fig.~\ref{fig:fidelity}, the success probability is calculated by taking the square of wave function overlap between the dynamical and targeted ground state. Again, at large $N$ the linear search algorithm fails to find the targeted state (the evolution time follows the  scaling of $T\propto\sqrt{N}$), while the RL-designed algorithm  still produces an adiabatic algorithm with high performance. The comparison to the nonlinear path is not shown here simply because for the hard Grover search Hamiltonian used here, the nonlinear path is not analytically available. 

In Fig.~\ref{fig:spectrum_HG} we plot the energy spectrum of the instantaneous Hamiltonian of the hard Grover problem for different numbers of qubits. The behavior of the ground and first excited states of the hard Grover problem is markedly different from those of the easy Grover case (Fig.~\ref{fig:combined_fig}). %As the number of qubit increases, the energy gap decreases significantly, which then makes adiabatic evolution much more difficult. This is due to the increased complexity of $H_B$ for the hard Grover case.
As can be seen from the plots, the linear protocol apparently starts to fail for $n= 6$.

%\medskip

\section{ Performance on 3-SAT}
%{\it Performance in 3-SAT.---} 
We  then apply the RL approach to the more complicated 3-SAT problem. Given a total number of $N_b$ boolean bits (labeled by $q$), the problem is to find a boolean sequence $z_q$ to satisfy 
$ 
C = C_1 \land C_2 \land C_3 \land C_4 \land\dots    
$
with each $C_i$ a clause containing three boolean bits, say $q_{i, k=1,2,3}$.  
The total clause number will be denoted as $N_C$. The satisfiability condition of each clause $C_i$ can be written into a truth table 
${\bf z}_{i\alpha} =\{ z^{(1)} , z ^{(2)} , z ^{(3)}  \} $ 
%\{ z_{q_{i, 1}}, z_{q_{i, 2}}, z_{q_{i, 3}} \} $
 such that the binary sequence $\{ z^{(1)} , z ^{(2)} , z ^{(3)} \} $ belongs to this table if and only if $C_i$ is satisfied. We use $\alpha$ to label all possibilities to satisfy the clause $C_i$.  To solve this problem with quantum adiabatic algorithm, we need to introduce $N_b$ qubits, which are then also labeled by $q$. The corresponding qubit states are $| z_{1}\rangle \otimes |z_{2} \rangle \otimes \ldots \otimes |z_{N_b}  \rangle$. Introducing a compact notation $|{\bf q}_i; {\bf z}_{i\alpha}\rangle$ for the qubits, ${q}_{i, 1} $, $q_{i, 2}$, and $q_{i, 3}$, in the quantum state $|z^{(1)}\rangle \otimes| z^{(2)}\rangle \otimes |  z^{(3)} \rangle$, the classical 3-SAT problem is formally encoded into a quantum ground state problem with a Hamiltonian 
\be 
\textstyle H_P ^{\rm SAT} = -\sum_{i=1}^{N_C}\sum_{\alpha}   | {\bf q}_i ; {\bf z} _{i\alpha}  \rangle \langle {\bf q}_i ;  {\bf z} _{i\alpha}  | . 
\ee 
A solution to the 3-SAT problem corresponds to a ground state of $H_p^{\rm SAT}$ with energy $-N_C$. Different 3-SAT problem instances correspond to different choices of clause ${\bf q}_i$ and truth table ${\bf z}_{i\alpha}$. 
The initial quantum state and Hamiltonian $H_B$ are set to be $|\psi_0\rangle$ and $H_B = \sum_q [\mathbb{1}-X_q]/2$, respectively, where $|\psi_0\rangle$ is the same initial state as in the Grover search problem.
In our RL approach to design 3-SAT quantum algorithm, the reward  RL-agent collects is generated by randomly sampling ${\bf q}_i$ and ${\bf z}_{i\alpha}$ (see Appendix~\ref{appendix:sampling}), to make the learned algorithm generically applicable.

%We first benchmark the RL-designed 3-SAT algorithm for the simplest problem, namely with one clause and three bits. 
%The results are shown in Fig.~\ref{fig:3fig_combine}\textbf{a}, which corresponds to a particular 3-SAT problem with all binary sequences of three bits except $\{ 001\}$ and $\{0 11\} $ belonging to the truth table. 
%For a fixed adiabatic time $T=?$, the success probability to reach the correct solution by the RL-designed algorithm is significantly higher than the conventional linear algorithm~\cite{farhi_quantum_2000}. The linear algorithm could incorrectly output $\{ 0 0 1 \}$ or $\{0 1 1\}$ with a total probability around $20\%$. In sharp contrast, this failure probability from RL-designed algorithm is {\bf below $2\%$??.}   Consequently, to achieve the same level of success probability, the required computational time $T$ by the RL-designed algorithm is significantly shorter than the linear algorithm. 

In Fig.~\ref{fig:3fig_combine}, we show the performance of the RL-designed algorithm and compare with the linear algorithm. We put the RL-agent to work on a $10$-bit 3-SAT problem. The RL-designed algorithm is obtained by training with clause number $N_C = 3$ only, where the stepwise reward is obtained by averaging over $100$ random problem instances. We then test the RL algorithm on random 3-SAT problem instances that contain one-to-six clauses. The tested success probability in Fig.~\ref{fig:3fig_combine}({\bf a}) is obtained by averaging over $10^5$ random problem instances. It is evident that the RL-designed algorithm outperforms the linear algorithm with higher success probability.  Its advantage  becomes more significant in a systematic fashion as the clause number is increased, although the RL algorithm is trained on 3-SAT problems with clause number $N_C = 3$ only. This implies the emergent transferability of the RL-designed algorithm. 
%It is worth emphasizing here that the RL-designed algorithm is obtained by training on 3-SAT problems with $3$ bits and $1$ clause only. 
The success  over different clause numbers implies that this RL-learning approach has seized the intrinsic ingredients  to optimize the adiabatic quantum  algorithm because otherwise the RL-designed algorithm would not be transferable.

Besides the quantitative improvement in the RL-designed over the linear algorithm, we also emphasize that the outcome of the RL algorithm is qualitatively distinct in the resultant fidelity. In Fig.~\ref{fig:3fig_combine}({\bf b},  {\bf c}), we show the distribution of the infidelity obtained from $10^5$ random 3-SAT problem instances. The statistics is taken for different clause number separately. The infidelity is rescaled by taking its average as a unit. The distributions of this rescaled infidelity for different clause numbers are found to collapse onto a universal function, for both the RL (Fig.~\ref{fig:3fig_combine}({\bf b})) and linear (Fig.~\ref{fig:3fig_combine}({\bf c})) algorithms. It appears that infidelity distribution from the linear algorithm is close to a Wigner-Dyson (WD) distribution---the numerically obtained statistical second moment of the infidelity agrees with the  WD prediction within $10\%$ difference.  
To the contrast, the second moment of the infidelity from the RL algorithm deviates from the WD prediction, meaning the infidelity distribution for the RL algorithm is qualitatively distinctive from the linear case. 
 The physical implication of such qualitative difference in the infidelity distribution is left for future study.

\begin{figure*}
\centerline{\includegraphics[width=.65\textwidth]{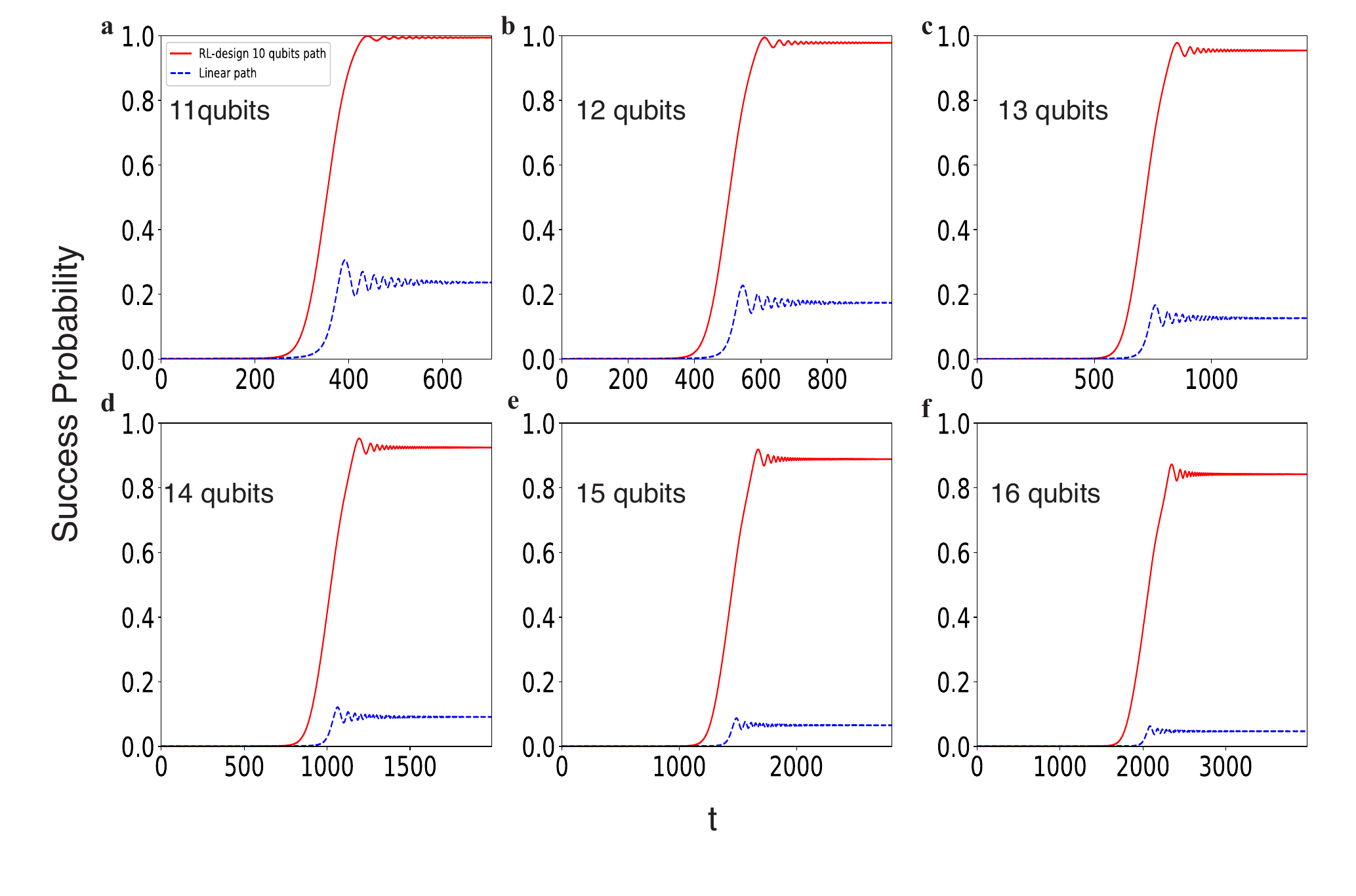}}
\vspace{-.5cm}
\caption{Performance of linear and  machine learning designed  quantum adiabatic algorithms applied to Grover search with different number of qubits. In this plot, the reinforcement learning (RL) designed schedule $s(t/T)$ is obtained by training on the problem with qubit number $n=10$, and then applied to problems having larger number of qubits, $11$ to $16$, by a simple rescaling $T\propto \sqrt{2^n}$. The RL-designed adiabatic algorithm systematically out-performs the linear one.}
\label{fig:transfer_learning}
\end{figure*}

\section{Scalability of the reinforcement learning in quantum adiabatic algorithm design on Grover Search}
\label{section:Scalability}
In Fig.~\ref{fig:transfer_learning} we show the results of applying a schedule learned on a $10$-qubit easy Grover search problem to $n$-qubit problems with $n>10$. For the linear algorithm, the schedule is $s(t/T) = t/T$, and $T$ scales according to $\sqrt{2^n}$ as we increase the qubit number in Fig.~\ref{fig:transfer_learning}. For the RL-designed algorithm, the schedule $s(t/T)$ is obtained by a training process on a problem with qubit number $n=10$, and then the schedule is applied to problems having larger number of qubits ($n = 11,\ldots, 16$), following the same rescaling $T\propto \sqrt{2^n}$ as the linear case. While the fidelity decreases as we increase the qubit number, it is evident that the RL-designed algorithm systematically out-performs the linear algorithm despite the simple rescaling applied. The comparison in the infidelity is explicitly given in Fig.~\ref{fig:transfer_infidelity}. 

To further demonstrate the scalability of the RL-learning, we provide the infidelity of the RL-designed algorithm trained on Grover search with qubit number $n$, and then applied successively on a problem with $n +1$ (see Fig.~\ref{fig:transfer_infidelity}). 
We use the number of training steps to quantify the resources spent on the RL-learning.  
Assuming access to an actual quantum computer, the number of training steps multiplied by the parameter $MI$ (see Section~\ref{sec:IIA} and Table~\ref{tab:paralist}) would be equal to the number of running times of the quantum computer.  
 
We emphasize here that the number of  training steps for different qubit number $n$ is fixed (see parameters of annealing protocol iteration $L_{SA}$ and path state iteration $L_{PS}$ in Table~\ref{tab:paralist}).
 The resultant infidelity in this iterative procedure remains close to $1\%$.  This further implies the schedule trained on relatively smaller-size problems has a rather large degree of transferability. 

In order to explicitly show that the reinforcement learning is helpful in obtaining a new schedule from a prior guess as we alter the problem, we provide the infidelity during the training process in Fig.~\ref{fig:pre_training}. We take the qubit number $n = 11$, and  compare two cases with and without pre-knowledge of the schedule. In Fig.~\ref{fig:pre_training} (a), the training process starts from a trivial linear schedule $s(t/T) = t/T$, i.e., no pre-knowledge given, whereas in Fig.~\ref{fig:pre_training} (b), the iteration starts from a Q-table already obtained through training on the problem with  $n=10$, i.e., with pre-knowledge.  It is evident that the reinforcement learning is indeed substantially helpful in quickly finding a new schedule from a prior guess even when the problem is altered---here the qubit number is changed from $n=10$ to $11$. With pre-knowledge, our reinforcement learning is able to find a proper schedule with three times smaller of iteration steps. 

\begin{figure}[htp]
\includegraphics[width=0.45\textwidth]{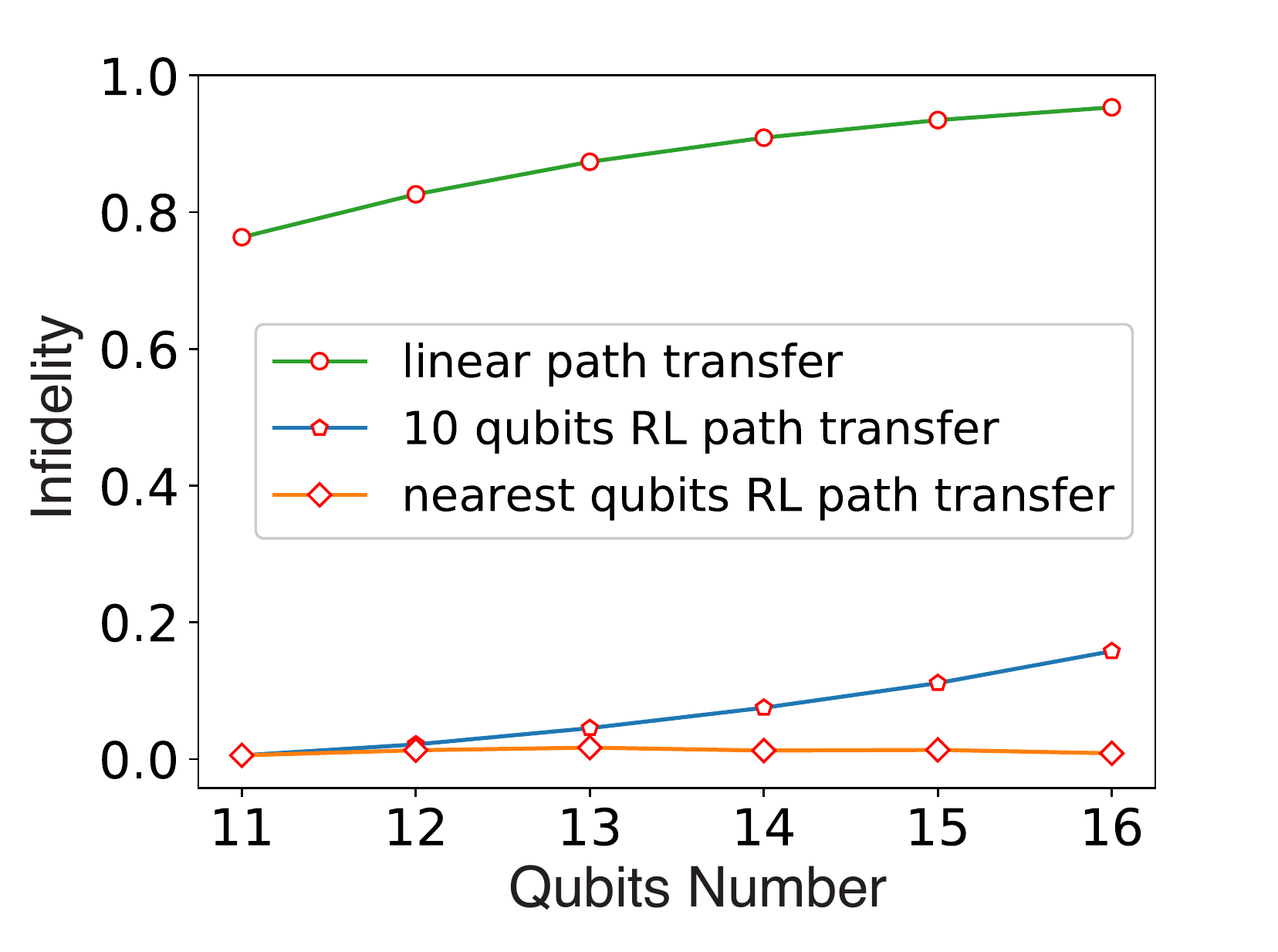}
\caption{Transferability of reinforcement learning designed schedules for Grover search. The `green' line shows the resultant  infidelity from the linear schedule $s(t/T) = t/T$ as a comparison.  The `blue' line shows the infidelity following the schedule that is obtained by training on the problem with qubit number $n=10$. The `orange' line corresponds to the schedules obtained by training on the problem with qubit number $n$ under a fixed number of training steps (see parameters of annealing protocol iteration $L_{SA}$ and path state iteration $L_{PS}$ in Table~\ref{tab:paralist} ) and then applied to the problem with qubit number $n+1$. In this plot, the total adiabatic time $T$ is chosen to scale with the qubit number as $T\propto \sqrt{2^n}$. 
  }
\label{fig:transfer_infidelity}
\end{figure}

\begin{figure*}
\centerline{\includegraphics[width=.8\textwidth]{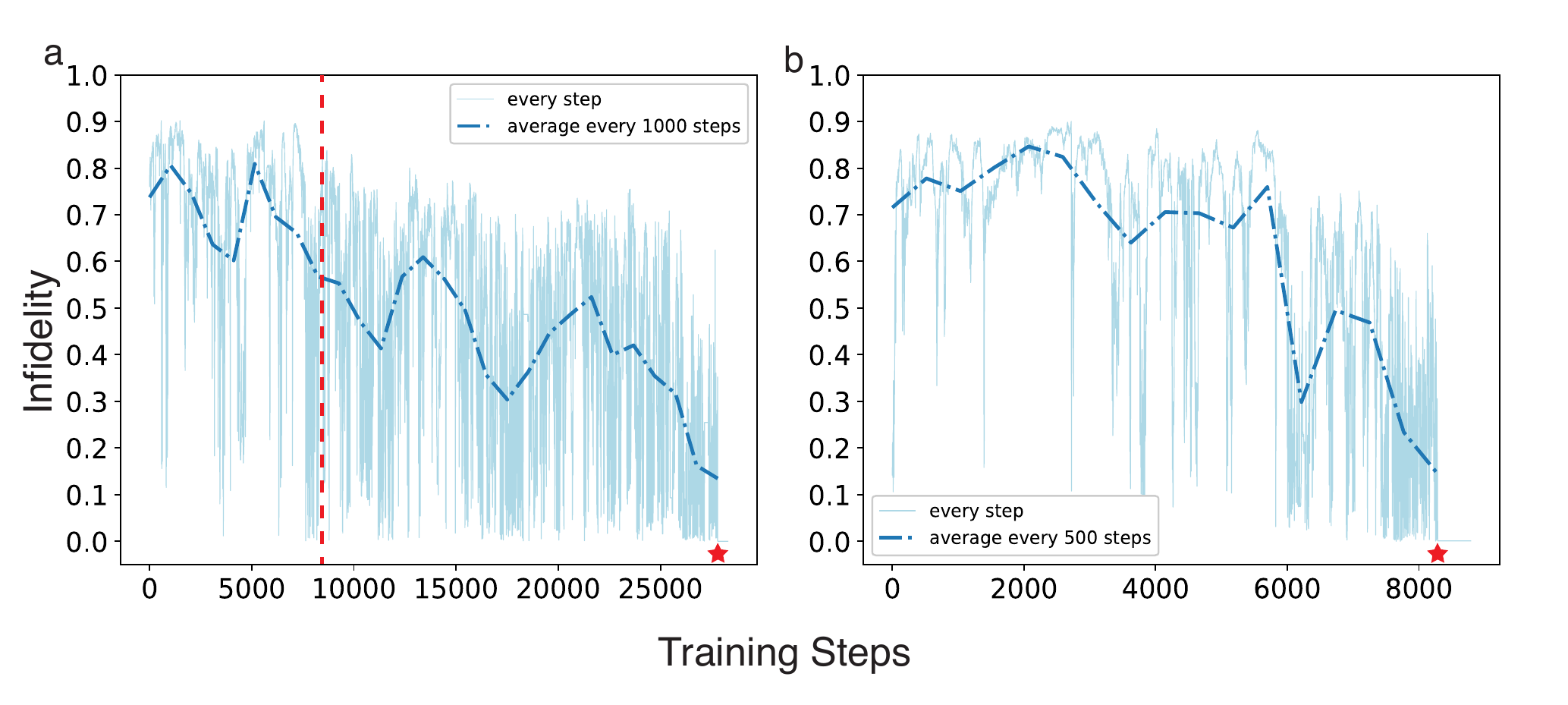}}
\caption{Stepwise infidelity during the reinforcement learning process (a) without and (b) with pre-knowledge.   The results in this plot correspond to easy Grover search problem with qubit number $n = 11$ In (a), the training process starts from a trivial linear schedule. In (b), the process starts from a schedule obtained for Grover search with qubit number $n= 10$ (rescaled according to $\sqrt{2^n}$). In both cases, the infidelity does not asymptotically converge to $0$ during the training process, but instead the chance for the learning agent to find a low-infidelity schedule is getting more frequent with more iteration steps. The red star in both plots marks the point where the performance reaches the threshold and $\epsilon_{max}$ in the $\epsilon$-greedy policy is set to be $1$. 
%The steps after the red star in both cases correspond to the $\epsilon$ greedy policy with increasing $\epsilon$ to $1$.} 
Comparing (a) and (b), at an iteration step around $8000$ (at the position of `red' vertical dashed line in (a) and the end of (b), the chance for the learning agent with pre-knowledge to find a low-infidelity schedule is substantially more frequent than that without pre-knowledge. The averaged infidelity in (b) drops down with much less iteration steps than in (a). 
 }
\label{fig:pre_training}
\end{figure*}

%\medskip 
%\ncsection{Summary and Discussion}
\section{Conclusion}
%{\it Conclusion.---} 
In this work we report a reinforcement-learning-based approach for automated quantum adiabatic algorithm design. 
Our devised approach is directly applicable to problems with solutions easy-to-verify such as searching, factorization, and NP-complete problems. 
%Further combined with circuit-based methods such as quantum phase estimation, this approach can be made generically applicable. 
Through numerical simulations, we show that the RL approach automatically finds an adiabatic algorithm for Grover search with quadratic speedup. 
%apparent advantage of the RL approach over  the conventional quantum adiabatic algorithms in Grover search and 3-SAT problems. 
In the application to the 3-SAT problems, we find  surprising transferability of the RL-designed algorithm which suggests the algorithm trained on  relatively-smaller size problems is applicable to larger sizes, which is both practically useful and theoretically inspiring in considering the complexity scaling. 
%We expect this aspect can be further improved by combining with transfer learning techniques. 
%The performance of RL enabled automated adiabatic algorithm design is expected to be systematically improvable by using more resources for training on a larger set of problem instances.  
The performance of our approach can be further improved by introducing additional Hamiltonian terms, which would easily fit into the framework proposed here.

\section{Acknowledgement}
%{\it Acknowledgement.---}
J.L. acknowledges helpful discussion with  Xiuzhe Luo. 
This work is supported by National Program on Key Basic Research Project of China under Grant No. 2017YFA0304204, National Natural Science Foundation of China under Grants No. 11774067, 11934002,  and Natural Science Foundation of Shanghai City (Grant No. 19ZR1471500), Shanghai Municipal Science and Technology Major Project (Grant No.2019SHZDZX04). 
XL would like to thank Department of Physics at Harvard University for hospitality during the completion of this work. 
The first two authors J.L. and Z.Y.L. contribute equally to this work.

\appendix
\begin{center} 
{\bf Appendix} 
\end{center}

\section{pseudo code and parameters of reinforcement learning }

For completeness, the pseudo code and the parameters used in our RL architecture for Grover search and 3-SAT problems are shown in algorithm~\ref{alg:architecture} and Table~\ref{tab:paralist}.

\begin{algorithm}[H]
  \caption{RL architecture for automated adiabatic quantum algorithm design.}
  \label{alg:architecture}
  \begin{algorithmic}[1]
    \INITIAL
      Initialize ``temperature" in simulation annealing  $Tem_0$;
      Initialize memory ${\cal M}$;
      Initialize path state;
      Initialize predict action-value $Q$ neural network with random weight $\theta$;
      Initialize target action-value $Q$ neural network with weight $\theta_{-} = \theta$;
    \FOR{ simulation annealing iteration $j=1,L_{SA}$}
      \STATE Set ``temperature" $Tem_{j} = Tem_{j-1}*10^{-C_R}$
        \FOR{path state iteration $i = 1,L_{PS}$}
    \STATE With probability $1-\epsilon$ select a random action $a_i$,and with probability $\epsilon$ select $a_i = argmax_a Q(\mathbf{b}_i,a;\theta)$. 
    \STATE Accept and execute action $a_i$ on $\mathbf{b}_i$ with probability  $P(e, Tem_j)$
    \STATE Get the next path state $a_i(\mathbf{b}_{i})$ and the reward $r_i$ by averaging the performance of $MI$ instances. 
    (The number of training steps multiplied by $MI$ is then equal to the number of running times of the quantum computer, which is simulated in our work. The number of training steps is thus a valid measure of computation resources spent on the RL-learning.)
    \STATE Store transition $(\mathbf{b}_i,a_i,r_i,a_i(\mathbf{b}_{i}))$ in memory ${\cal M}$
    \STATE Sample a batch of transitions$(\mathbf{b},a,r,a(\mathbf{b}))$ from memory randomly
    \STATE Perform a stochastic gradient descent on the loss function $L(\theta )= \mathbb{E} _{\mathbf{b},a,r,a(\mathbf{b})}[(r+\gamma \max_{a'} Q(a(\mathbf{b}),a';\theta_{-}))-Q(\mathbf{b},a;\theta)]^2$ with respect to the predict network parameter $\theta$
    \STATE Increase $\epsilon$ with single step increment. Here set upper bound of $\epsilon$ value $\epsilon_{max}$
    \STATE Every $W$ steps, set $\theta_{-} = \theta$
    \IF {$r_i \geq threshold$}
    \STATE Keep on path state updating iteration without training and set $\epsilon_{max}$ to be $1$
    \ENDIF
      \ENDFOR

    \ENDFOR
    \RETURN path state
  \end{algorithmic}
\end{algorithm}

\label{appendix:parameters}
\begin{table*}[thp]
	\centering
	\caption{Parameter list in the reinforcement learning}
	\begin{tabularx}{\textwidth}{*{3}{>{\centering\arraybackslash}X}}
		\hline  
		\diagbox{Parameters}{Problems}&Easy/Hard Grover Search&3-SAT\\ 
		\hline 
		Neural-network layer number			&2&3\\
		Neural-network hidden-layer neurons 	& 20 &12\\
		Neural-network learning rate			&0.01&0.01\\
		Neural-network activation function 		&relu&relu\\
		Training bath size					&32&32\\
		Reward discount factor,  $\gamma$		&0.9&0.9\\
		Memory capacity, {\it CAP} &500&1000\\
		Maximal $\varepsilon$-value in $\epsilon$-greedy policy, $\epsilon_{max}$	
										&0.9&0.9\\
		%The $\varepsilon$-value  in  final step of agent exploration &1.0&1.0\\
		Single-step increment of $\epsilon$ 		&0.01&0.01\\
		Target-Net refreshing parameter, $W$	&50&50\\
		Cooling rate in annealing protocol, $C_R$
										&0.1&0.1\\
		Initial ``temperature" in annealing, $Tem_0$				
										&10&10\\	
		Cutoff, $C$						&6&6 \\ 
		Maximal update per step, $\Delta_0$		&0.1&0.1 \\ 
		Problem instance averaging number, {\it MI}	
										&1&100 \\ 
		Annealing protocol iteration, $L_{SA}$		&80&80 \\
		Path state iteration, $L_{PS}$		&1000&1000 \\
		%Time Step $dt$&0.1&0.1\\
		\hline  
	\end{tabularx}
\label{tab:paralist} 
\end{table*}

\section{Sampling of 3-SAT problem instances }
\label{appendix:sampling}
Given a total number $N_b$ of boolean bits $z_q$, different 3-SAT problem instances correspond to different choices of three-bit combinations ${\bf q}_i$ in each clause $C_i$, and different choices of the truth table  of each clause defined to be ${\bf z}_{i\alpha} = (z^{(1)} , z^{(2)}, z^{(3)})$ in the main text. Since we aim at a quantum adiabatic algorithm generically applicable, we randomly sample the problem instances $\{ {\bf q}_i, {\bf z}_i \}$ according to the definition of 3-SAT problem. 
%and take the averaged energy of the final quantum state of the adiabatic evolution to assign the reward of a path-state ${\bf b}$. 
It is worth noting here that the choice for the truth table is not completely random, and that for one clause in the 3-SAT problem, there are eight possibilities of choosing the truth table corresponding to the eight possibilities of constructing the clause. The size of the sampling space grows polynomially with $N_b$ and exponentially with $N_C$.

\bibliography{references}

\end{document}